\documentclass[fleqn,usenatbib]{mnras}

\usepackage{newtxtext,newtxmath}

\usepackage[T1]{fontenc}
\usepackage{ae,aecompl}

\usepackage{graphicx}	
\usepackage{amsmath}	
\usepackage{amssymb}	
\usepackage{lineno}
\DeclareGraphicsExtensions{.png,.pdf,.jpg}
\usepackage{color}
\usepackage{ragged2e}

\newcommand{\ic}{IC~1531}
\newcommand{\Cha}{{\it Chandra}}
\newcommand{\swift}{{\it Swift}}
\newcommand{\xmm}{{\it XMM-Newton}}
\newcommand{\ergcm}{erg cm$^{-2}$ s$^{-1}$}

\newcommand{\dg}{$^\circ$}
\newcommand{\er}{$\pm$}

\def\mnras{MNRAS}
\def\apj{ApJ}
\def\apjl{ApJL}
\def\apjs{ApJS}
\def\aap{A\&A}

\def\aapr{A\&A Rev.}
\def\araa{ARA\&A}
\def\aj{AJ}

\def\physrep{Phys. Rep.}
\def\pasp{Publ. Astr. Soc. Pac.}

\def\pasj{Pub. Astron. Soc. Japan}
\def\memsai{Mem. Soc. Astron. Ita.}

\newcommand{\lumunits}{erg cm$^{-2}$ s$^{-1}$}

\newcommand{\g}{$\gamma$}

\newcommand{\fermi}{${\it Fermi}$}
\newcommand{\ergs}{erg s$^{-1}$}

\title[IC~1531]{Faint $\gamma$-ray sources at low-redshift: the radio galaxy IC~1531}

\author[Bassi T. et al.]{
T. Bassi,$^{1,2,3}$\thanks{E-mail: bassi@ifc.inaf.it}
G. Migliori,$^{4,5,6}$
P. Grandi,$^{7}$
C. Vignali,$^{4,7}$ 
M.~A. P{\'e}rez-Torres,$^{8}$
\newauthor
R.~D. Baldi,$^{9}$
E. Torresi,$^{4,7}$
A. Siemiginowska,$^{10}$
C. Stanghellini$^{5}$\\
$^{1}$ INAF -- Istituto di Astrofisica Spaziale e Fisica Cosmica di Palermo, Via Ugo La Malfa 153, 90146 Palermo, Italy \\
$^{2}$ Universit\'a degli Studi di Palermo, Dipartimento di Fisica e Chimica, via Archirafi 36 - 90123 Palermo, Italy\\
$^{3}$  IRAP, Universit\'{e} de Toulouse, CNRS, UPS, CNES, Toulouse, France\\
$^{4}$ Dipartimento di Fisica e Astronomia, Alma Mater Studiorum, Universit\`a degli Studi di Bologna, Via Gobetti 93/2, 40129 Bologna, Italy \\
$^{5}$ INAF Istituto di Radioastronomia, via Gobetti 101, 40129 Bologna, Italy \\
$^{6}$ Laboratoire  AIM  (CEA/IRFU  -  CNRS/INSU  -  Universit\'{e}  Paris  Diderot),  CEA  DSM/SAp,  F-91191  Gif-sur-Yvette, France \\
$^{7}$ INAF -- Osservatorio di Astrofisica e Scienza dello Spazio di Bologna, Via Gobetti 93/3, 40129 Bologna, Italy \\
$^{8}$ Instituto de Astrof\'isica de Andaluc\'ia (IAA-CSIC), E-18008 Granada, Spain \\
$^{9}$ School of Physics and Astronomy, University of Southampton, Southampton, SO17 1BJ, UK \\
$^{10}$ Harvard Smithsonian Center for Astrophysics, 60 Garden St, Cambridge, MA 02138, USA}

\date{Accepted XXX. Received YYY; in original form ZZZ}

\pubyear{2017}

\begin{document}
\label{firstpage}
\pagerange{\pageref{firstpage}--\pageref{lastpage}}
\maketitle

\begin{abstract}
We present a multi-wavelength study of \ic~(z=0.02564), an extragalactic radio source associated with the \g-ray object 3FGL J0009.9$-$3206 and classified as a blazar of uncertain type in the Third \fermi~Large Area Telescope AGN Catalog (3LAC). A core-jet structure, visible in radio and X-rays, is enclosed within a $\sim$220 kpc wide radio structure.
The morphology and spectral characteristics of the kiloparsec jet in radio and X-rays are typical of Fanaroff-Riley type I galaxies. The analysis of the radio data and optical spectrum and different diagnostic methods based on the optical, infrared and \g-ray luminosities also support a classification as a low-power radio galaxy seen at moderate angles ($\theta=$10\dg--20\dg). 
In the framework of leptonic models, the high-energy peak of the non-thermal nuclear spectral energy distribution can be explained in terms of synchrotron-self-Compton emission from a jet seen at 
 $\theta\sim$15\dg. 
Similarly to other misaligned AGNs detected by \fermi, 
the required bulk motion is lower ($\Gamma_{\rm bulk}=$4) than the values inferred in BL Lac objects, confirming that, because of the de-boosting of  emission from the highly-relativistic blazar region, these nearby systems are valuable targets to probe the existence of multiple sites of production of the most energetic emission in the jets. 
\end{abstract}

\begin{keywords}
galaxies: active -- galaxies: individual: \ic~-- galaxies: jets -- gamma-rays: galaxies -- radio continuum: galaxies.
\end{keywords}



\section{Introduction}
A decade of observations of the Large Area Telescope (LAT) on board the {\it Fermi} satellite has significantly expanded our knowledge of the $\gamma$-ray sky. Even though the extra-galactic $\gamma$-ray sky remains firmly dominated by blazars \citep{fermi}, i.e. active galactic nuclei (AGN) with a jet pointing toward the observer, we have for the first time the possibility to probe the population of faint $\gamma$-ray sources. This is comprised of luminous sources at high-redshift or intrinsically faint sources at low redshift. The former are important to understand the cosmological evolution of $\gamma$-ray blazars \citep[][and references therein]{Aje12,Aje14}, hence to trace the growth of the super-massive black holes (SMBHs) powering the extragalactic jets \citep{Vol11,Ghi13,Ack17}. The latter includes low-luminosity blazars \citep{Mas17} and a heterogeneous ensemble of sources. 

Non-blazar faint $\gamma$-ray sources are misaligned AGNs \citep[MAGNs,][]{Abdo} and star-forming galaxies \citep{Ack12}. In addition, there is a large number of unidentified sources, which hold potential for the discovery of new classes of $\gamma$-ray emitters \citep[see e.g.  the class of narrow-line Seyfert galaxies, the class of FR~0s \citep{baldi15}, Tol~1326$-$379, PKS~1718$-$649,][]{Fos11,Dam16,grandi16,Mig14,Mig16,tavecchio18}.
The LAT MAGNs, i.e. AGNs with a jet that is seen at moderate/large angles, collect radio galaxies (RGs) and steep-spectrum radio quasars (SSRQs). 
According to the unified model of AGN, MAGNs are the misaligned counterparts of blazars. In blazars  the detection of the high-energy emission is favored by relativistic effects, which boost and blue-shift the emission from a compact region (so-called blazar region) of the jet moving at high Lorentz factors \citep[$\Gamma\sim$10--20,][]{Celotti08}. 
Since the emission from the blazar region is de-boosted and red-shifted at increasing viewing angles, the detection of MAGNs in the GeV band may be indicative of a different emission site or emission mechanism.
An example is the radio galaxy Centaurus A, where the extended $\gamma$-ray component is most likely associated with the radio lobes \citep{CenA10,Sun16}. \\
\indent
Modeling of the spectral energy distributions (SEDs) of the LAT MAGNs with one-zone leptonic models implies relatively slow jets with respect to blazars \citep{Chiaberge}.
Models of stratified jets, where the emission from the fast spine is visible at small viewing angles and that from the slow moving layer emerges at larger angles, have been considered as a viable alternative \citep{GK03,spine-layer}. \\
\indent
In order to test the different scenarios, we need to target the population of faint $\gamma$-ray sources at low redshift. In particular, sources that mark the transition between MAGN and blazars may help to establish whether there is a continuity between the two populations, with the inclination angle as the main parameter driving the change in the SED.
Understanding the link between orientation and $\gamma$-ray emission is also important to constrain the contribution of MAGNs to the extragalactic $\gamma$-ray background (EGB). In fact, according to recent studies \citep{Ino11,DiM14,For15}, radio galaxies and star-forming galaxies could account for a significant fraction of the EGB.\\
\indent
In this paper, we present a multi-wavelength study of the radio source \ic~ (z=0.02564), which is associated with the LAT source 3FGL~J0009.9$-$3206.
In the third Catalog of AGN detected by \fermi~LAT \citep[3LAC,][]{Acker} it is classified as a blazar of  uncertain type.
The main goals of our study are to assess the classification of \ic~and to investigate the nature of its high-energy emission.
Throughout this paper, the following cosmological parameters are assumed: H$_{0}$=67.8 km s$^{-1}$ Mpc$^{-1}$, $\Omega_{m}=0.3$ and $\Omega_{\lambda}$=0.7 \citep{Pla16}. At the source distance, 1" corresponds to 514 pc.

\section{IC~1531}\label{source}
\ic~(also known with the name PKS 0007-325) is an early-type lenticular galaxy \citep{loveday} located in  a low-density environment \citep{Sullivan}. It is present in several surveys and catalogs but it has not been extensively studied.

\begin{figure}
	\includegraphics[width=\columnwidth]{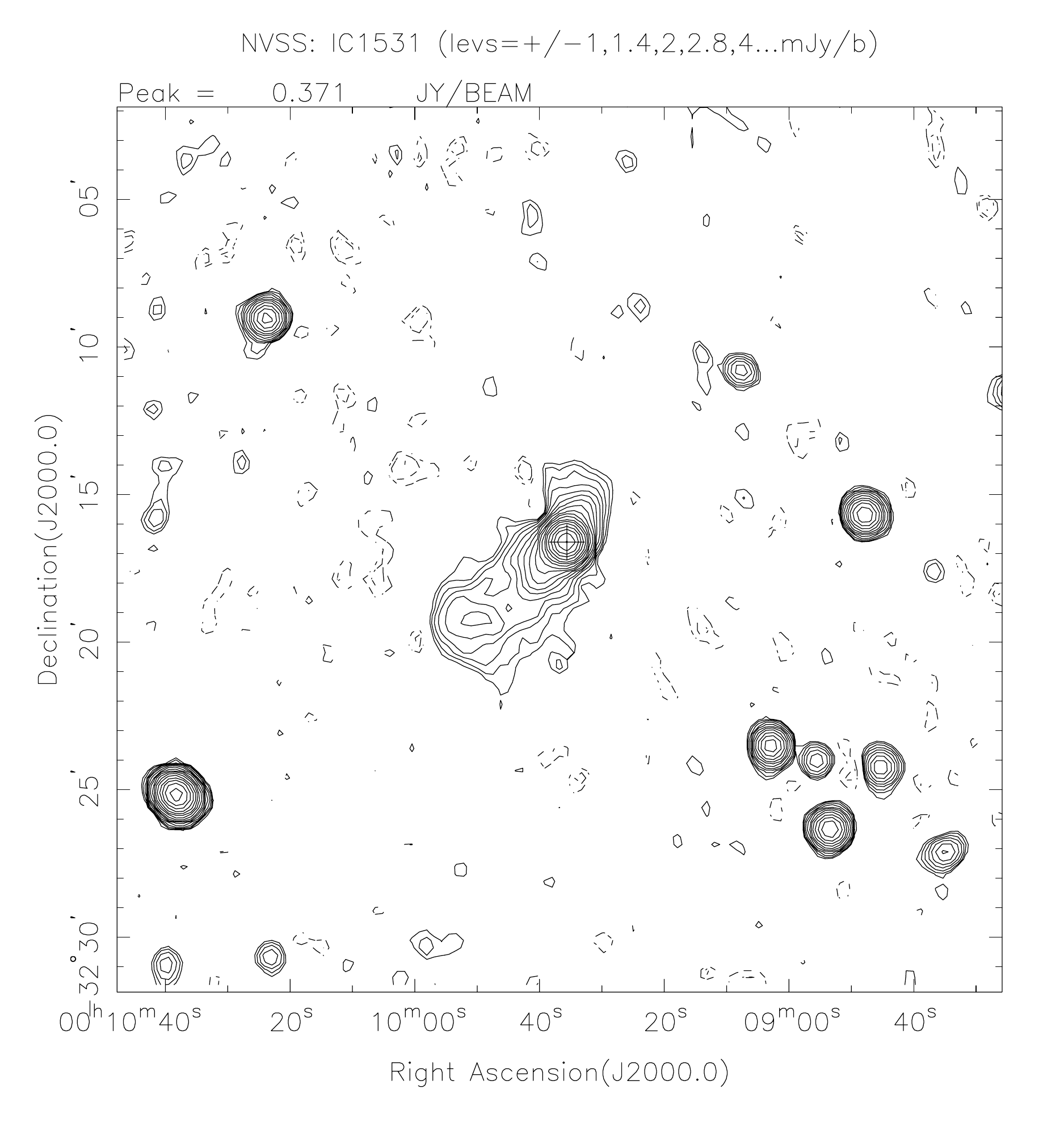}
    \caption{NVSS contour map of IC~1531 at 1.4 GHz. The radio structure in the center of the image is $\sim$220 kpc extended.}
    \label{NVSS}
\end{figure}

The radio map at 1.4 GHz of the {\it National Radio Astronomy Observatory (NRAO) Very Large Array Sky Survey} \citep[NVSS,][]{condon98} shows an asymmetric jetted radio source (see Fig. \ref{NVSS}), which is powered by the active nucleus. The south-eastern radio lobe has a $\sim$220 kpc projected extension. The total flux density at 1.4 GHz is 350 mJy \citep{ekers}. In Sec. \ref{radio} we analyzed the archival radio observations and obtained the flux densities of the radio components at 1.4--1.5 and 8.4 GHz (see Table \ref{sed}).

The measured flux densities near 1 mm lie close to the extrapolation of the core data at radio frequencies, supporting a non-thermal origin of the millimeter flux, while the emission of the host galaxy clearly emerges in the infrared-optical band \citep{Knapp}. 
The weak excess that is visible at 100 $\mu$m is most likely of thermal origin and due to radiation from cool dust \citep{Knappaltro} heated by massive stars \citep{kennicutt}. 

Based on the optical spectrum reported in the {\it 6 degree Field Galaxy Survey} \citep[6dFGS,][]{jones04,jones09}, IC~1531 was classified as an AGN with weak lines in absorption and emission \citep{Mahony}.

\citet{Sullivan} studied the galaxy and its environment in X-rays using observations by the \xmm~mission and the {\it Chandra X-ray observatory}. The central AGN dominates the X-ray output ($F_{[0.4-7]\;keV}\sim$4.2$\times 10^{-13}$ \ergcm), together with an extended ($\lesssim$7.5\arcsec) component oriented in the direction of the radio jet \citep{Sullivan}. 
The source is surrounded by a relatively cold ($\sim$0.55 keV) and compact X-ray halo \citep{Sullivan}. The authors concluded that the AGN could have induced gas-mass loss either via AGN-driven winds or via jet stripping.

In \g-rays, \ic~is associated with 3FGL J0009.9-3206, a source reported in the {\it third Fermi Large Area Telescope source catalog } \citep[3FGL,][]{fermi}.
3FGL J0009.9-3206 is detected at the 5.5$\sigma$ significance level with a flux in the 1-100 GeV band of $(2.77\pm 0.62)\times 10^{-10}$ $ph\;cm^{-2}\;s^{-1}$ and a power law photon index  $\Gamma_{\gamma}=$2.3$\pm$0.1.

\section{Radio}\label{radio}

We downloaded publicly available Karl G. Jansky Very Large Array (VLA) data of IC~1531 taken 
in A configuration at 1.5 GHz on 7 Jul 1987, at 1.4 GHz on 24 Jul 1995 (project codes AV0151 
and AG0454, respectively), and at 8.5 GHz on 19 May 1998 (project code AH0640). 
We used standard routines within the NRAO {\it AIPS} package for all data reduction steps, 
including amplitude and phase calibration,
as well as imaging; our results are summarized in Table \ref{VLA-log}. 
We show in Fig. \ref{IC 1531} the 8.5 GHz image of IC~1531 (contours), where 
the coincidence of the radio core with the X-ray nuclear emission is evident. 
The position of the 8.5 GHz peak is RA=00h09m35.56057s, Dec=$-$32 16 
36.9000 (J2000.0). 

The 1.4-1.5 GHz flux densities (both peak and total) are compatible with
a non-variability of the source in the two epochs, assuming a 10\% uncertainty in the amplitude
calibration. In fact, the differences in the flux densities of the peaks are
enhanced because of the different FWHM sizes. We therefore restored the two
images with a common beam. The difference in the resulting peak flux densities was, as expected, smaller than in the case of the natural synthesized beams, confirming
that there is neither evidence for nuclear, nor extended variability of the
source within 1-$\sigma$. 

\begin{table*}
\caption{Log of the VLA observations}
\begin{center}
\begin{tabular}{lcccccc}
\hline
\hline 
 Date & Frequency & FWHM & rms & $\sigma$ & Peak flux density & Total flux density \tabularnewline
  & (GHz) & (arcsec$^2$) & (mJy\,beam$^{-1}$) &  (mJy\,beam$^{-1}$) & (mJy\,beam$^{-1}$) & (mJy)\tabularnewline
 \hline
1987-Jul-07 & 1.47 & 3.6$\times$1.3 & 0.7 & 24 & 239 & 340 \tabularnewline
1995-Jul-24 & 1.38 & 2.4$\times$1.2 & 0.2 & 20 & 203 & 332 \tabularnewline
1998-May-19 & 8.46 & 0.6$\times$0.2 & 0.1 & 15 & 296 & 404 \tabularnewline 
\hline
\tabularnewline
\end{tabular}
\end{center}
\justifying
\vglue-0.1cm
The FWHM corresponds to the natural synthesized beam obtained at each epoch. The quoted 1-$\sigma$ values are the result of adding in quadrature the quoted rms value for each map and the calibration uncertainty, assumed to be of 10\% at 1.4-1.5 GHz and 5\% at 8.5 GHz.
\label{VLA-log}
\end{table*}

\begin{table*}
\caption{Log of the X-ray observations}
\begin{center}
\begin{tabular}{lcccr}
\hline
\hline 
Telescope/Instrument  &ObsID & Date        &Exp. time   &Net count rate\tabularnewline
 \hline
\xmm/pn               &0202190301      &2004-May-20  &12.5           &2.8$\times$10$^{-1}$  \tabularnewline
\xmm/MOS1               &0202190301      &2004-May-20  &20.5           &7.3$\times$10$^{-2}$  \tabularnewline
\xmm/MOS2               &0202190301      &2004-May-20  &20.5           &7.7$\times$10$^{-2}$  \tabularnewline
\Cha/ACIS-S           &5783      &2005-Aug-21  &39.5           &1.1$\times$10$^{-1}$  \tabularnewline
\swift/XRT            &0046394001      &2011-Nov-12  &1.0           &1.2$\times$10$^{-2}$  \tabularnewline
 \swift/XRT     &0046394002      &2012-Apr-22  &3.7           &1.7$\times$10$^{-2}$  \tabularnewline
\hline
\tabularnewline
\end{tabular}
\end{center}
\justifying
\vglue-0.1cm
The exposure times are corrected for bad time intervals (e.g. high flaring). The net count rates refer to the extraction region of the total emission of each instrument (see Sec. 4) in the 0.5-7 keV energy band.
\label{Xray-log}
\end{table*}

\section{X-ray Observations and Analysis}
\label{Xrays}
\begin{figure}
	\includegraphics[width=\columnwidth]{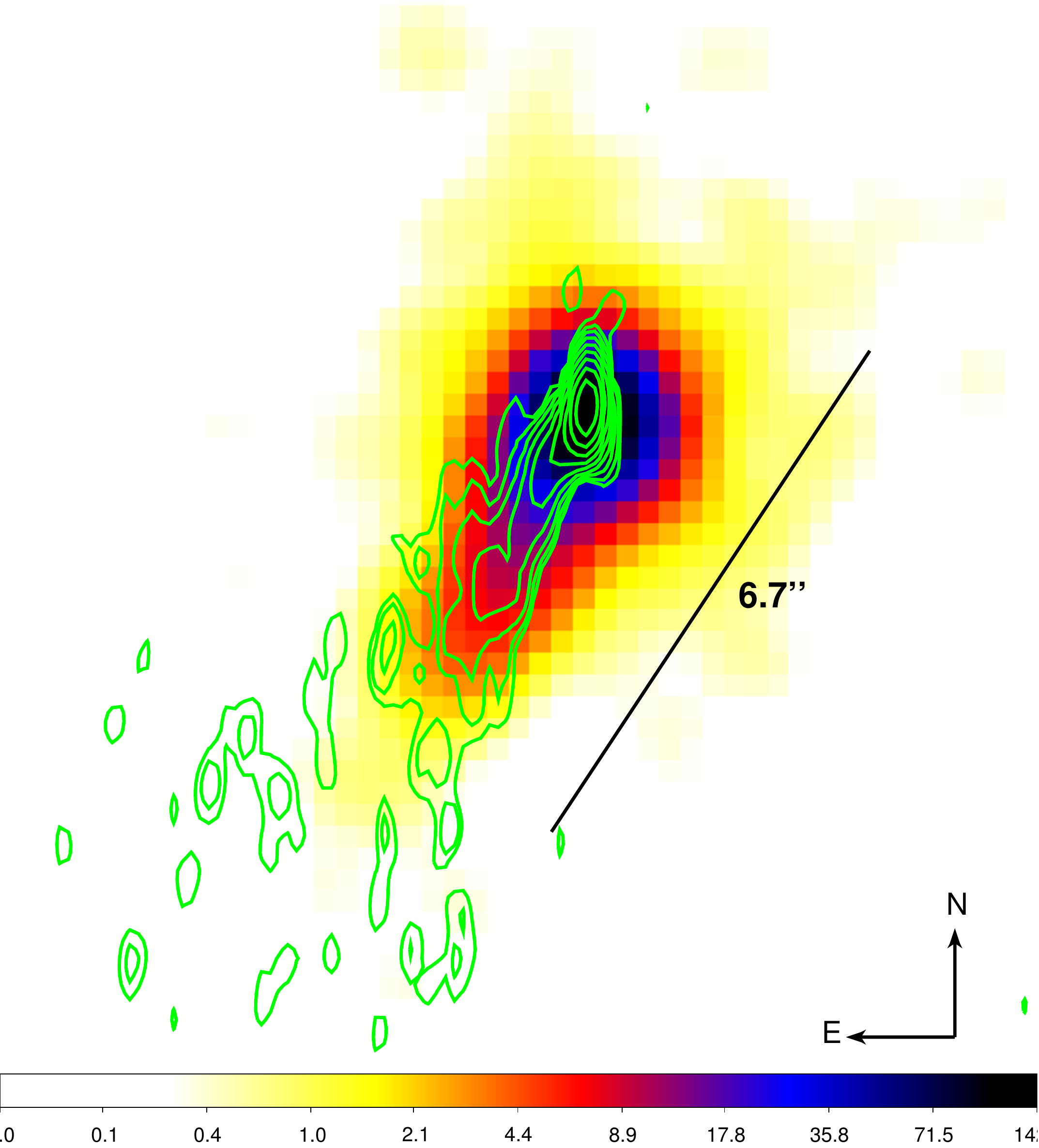}
    \caption{\Cha~0.5-7 keV ACIS-S image, at 0.492 arcsec/pixel resolution, and overlaid VLA 8.5 GHz contours of \ic. Contours level are 4rms$\times\sqrt{2}^n$, with rms$=$0.1 mJy beam$^{-1}$, the natural beam size is 0.6$\times$0.2 arcsec$^2$.}
    \label{IC 1531}
\end{figure}

We retrieved and analyzed the publicly available \Cha, \xmm~and \swift~observations of \ic~(see Tab. \ref{Xray-log}).\\
\ic~ was observed with the ACIS-S camera onboard the \textit{Chandra X-ray observatory} for $\sim$40 ks on 2005 August 21 (ObsID 5783).
We reprocessed the data using the \texttt{chandra$_{-}$repro} script in the \textit{Chandra Interactive Analysis of Observation (CIAO)} 4.7 software and the \textit{Chandra Calibration Database} CALDB v.4.6.9. We excluded high-flaring background periods. The net exposure time after filtering was 39.2 ks. We applied the sub-pixel event-repositioning algorithm (\texttt{pix$\_$adj=EDSER}), which optimizes the image resolution.\\
In Figure \ref{IC 1531}, we show the 0.5--7 keV \Cha~image overlaid the VLA radio contours at 8.4 GHz (see Section \ref{radio}). The unresolved X-ray core of \ic~is clearly visible. Extended (6.7\arcsec) X-ray emission is present in the region coincident with the south-eastern radio jet together with a faint X-ray halo around the point source.\\ 

We used \texttt{specextract} to obtain the spectra, and the relative response files, of the X-ray emission in the core and the jet regions. For the core, we selected a circular region of 1.4\arcsec~radius centered on the unresolved source. The spectrum of the extended jet was extracted from a rectangular box (angular dimensions of 4.2\arcsec$\times$5.4\arcsec). 
In addition, for a comparison with the \xmm~observation, we extracted the spectrum of the total X-ray emission of \ic~using an elliptical region (4.1\arcsec~and 6\arcsec~being the angular dimensions of the minor and major axis, respectively).
We extracted the background of the core from 5 contiguous circular regions (1.1\arcsec~radius) which are free from X-ray sources.
In order to apply the $\chi^2$  statistics, the data were grouped so that there are at least 15 counts in each spectral bin.
Using PIMMS, we estimated a 10\% pile-up fraction for the 0.5--7 keV count rate of 7.96 cts s$^{-2}$ in the core region.
The background spectrum of the extended jet was extracted from a region at the same distance as the jet (1.4\arcsec~to 6.8\arcsec) from the X-ray peak, in order to account for the possible contribution of the diffuse thermal emission and the core's non-thermal emission\footnote{In fact, up to 10\% of the point-source flux should be spread at $\gtrsim$1.4\arcsec~radius from the peak of the emission (see the Proposer's Observatory Guide, http://cxc.harvard.edu/proposer/POG/).}. The jet and background spectra were simultaneously fit. 
The background best-fit model was a steep power-law ($\Gamma=$2.9), and the estimated contribution to the 2-10 keV flux is below 10\%.
\\ 
\indent
A $\sim$24 ks \xmm~observation of \ic~was performed on May 20, 2004.  
The raw data were processed with the \xmm~Science Analysis System (SAS v 14). After filtering for background flares, the net exposure times were 12.5 ks for the pn and 20.5 ks for the MOS1 and MOS2. 
The X-ray emission of the core and jet of \ic~are unresolved at the angular resolution of \xmm~(FWHM $\sim$6\arcsec). Therefore, we extracted a single 0.5--7 keV spectrum from a circular region of 30\arcsec~radius for the pn and 40\arcsec~radius for the MOS1 and MOS2 (enclosing $\sim$90\% of the total flux of a point-like source) and verified that the emission is not affected by pile-up effects. The background spectra were obtained from blank-sky regions (30\arcsec~ and 40\arcsec~radius respectively for pn and MOS) located on the same chip of our source. 
We grouped the spectra so that each spectral bin has a minimum of 25 counts. 
\\
\indent
The \textit{X-Ray Telescope} (XRT), on board the {\it Neil Gehrels Swift Observatory} observed \ic~twice in photon-counting mode: the first observation (obsID=00046394001) was carried out on November 12, 2011 (1.04 ks), while the second pointing (obsID=00046394002) targeted the source on April 22, 2012 (3.72 ks).
The X-ray data were processed and spectra were extracted using the on-line product generator available at http://www.swift.ac.uk/user$\_$objects/ using default settings \citep{evans09}.
The number of net (i.e., background-subtracted) counts in the 0.3--10~keV band are 12 and 63; fitting the admittedly low counting statistics spectra separately with Cash statistics \citep{cash79} and adopting a power law model indicated that no significant variability was present between the two observations, so the two datasets were summed.
The final spectra were grouped to have at least 1 counts per spectral bin, so the data can be fitted with the Cash statistic.\\

\subsection{Spectral analysis results}
\label{Xres}
The \Cha~spectrum of the core region is best fitted by a composite model with a steep power law ($\Gamma=$2.2\er0.1) plus a thermal component (an APEC model in {\sc xspec}) with a plasma temperature of 0.6\er0.2 keV. The Galactic column density is N$\rm _H=1.69\times10^{20}$ cm$^{-2}$ \citep{Kal05}. 
The best-fit parameters are reported in Table \ref{Chandra-sp}.  \\
\indent
We observed a deviation of the data from the model in the soft (0.5--2 keV) energy band. In principle, the flattening of the spectrum could be due to an intrinsic absorber (N$\rm _{H,int}=4.50\times10^{20}$ cm$^{-2}$),  however this additional spectral component is not required to model the \xmm~ and \swift~spectra, suggesting that this is rather an effect of the pile-up.\\
The fit of the spectrum of the total emission of \ic~gave consistent results. Confirming the results of \citet{Sullivan}, the X-ray emission is dominated by the contribution of the AGN: the unabsorbed non thermal luminosity in the 0.5--10 keV energy band is 8.3$\times$10$^{41}$ \ergs.  
The thermal radiation of the gas halo (L=7.4$\times$10$^{40}$ \ergs) represents less than 10\% of the total X-ray luminosity.

A comparison between the fits of the \Cha~and \xmm~spectrum shows a slight difference in the fraction of the soft X-ray flux attributed to each, thermal and non-thermal, component. However, this seems rather due to a combination of effects, i.e. the different sizes of the extraction region and the higher sensitivity of \xmm~with respect to \Cha~to low surface brightness.
Similarly, although there is an indication of a decrease of the X-ray flux in the \swift~observations, the limited statistics of the \swift~spectra leaves the spectral parameters unconstrained and makes difficult to verify it. Thus, we conclude that, if real, a variation of the X-ray flux between 2004/5 and 2012 would be smaller than a factor of two. 

The X-ray emission of the extended jet is continuous and progressively decays with the increasing distance from the core. 
The best-fit model of the extended jet spectrum is again a power law with a steep  photon index ($\Gamma=$2.2\er0.2). The unabsorbed non-thermal X-ray flux is $F_{X,jet}=(4.2\pm0.6)\times 10^{-14}$ \ergcm, which corresponds to a 2--10~keV luminosity of $6.6\times 10^{40}$ \ergs.

\begin{table*}
\caption{Best-fitting model (power law$\times$APEC) of the X-ray spectrum extracted from \ic~ core region. The Galactic column density is N$\rm _H=1.69\times10^{20}$ cm$^{-2}$ \citep{Kal05}. The errors are quoted at 90\% level for each parameter \citep{avni76}.}
\begin{center}
\begin{tabular}{ccccccc}
\hline
\hline 
 kT& $\Gamma$ & $\chi_{\nu}^{2}$(dof) & F$_{\left[0.5-2\right]\,PL}$ & F$_{\left[0.5-2\right]\, TE}$ & F$_{\left[2-10\right]\, PL}$ & L$_{\left[2-10\right]\, PL}$\tabularnewline
(keV)& &&(\lumunits) & (\lumunits)& (\lumunits) &(\ergs)\tabularnewline
(1)&(2)&(3)&(4)&(5)&(6)&(7)\tabularnewline
\hline
\tabularnewline
$0.6_{-0.2}^{+0.2}$ & $2.2_{-0.1}^{+0.1}$ & 1.16(123) & $2.8_{-0.3}^{+0.3}$ & $0.2_{-0.1}^{+0.1}$ &$2.5_{-0.2}^{+0.3}$ & $3.9_{-0.4}^{+0.4}$\tabularnewline
\tabularnewline
\hline 
\tabularnewline
\end{tabular}
\end{center}
\justifying
Columns: 1- Plasma temperature (APEC component); 
2- photon index; 
3- reduced $\chi^{2}$; 
4- unabsorbed 0.5-2 keV flux of the power-law component in unit of 10$^{-13}$ \lumunits; 
5- unabsorbed 0.5-2 keV fluxes of the thermal component in unit of 10$^{-13}$ \lumunits;
6- unabsorbed 2-10 keV flux of the power-law component in unit of 10$^{-13}$ \lumunits;
7- unabsorbed non-thermal 2-10 keV luminosity in units of 10$^{41}$ \ergs.
\label{Chandra-sp}
\end{table*}

\section{Optical properties}\label{opt}

IC~1531 has been classified as an AGN with weak narrow emission lines \citep{Mahony}. 
In fact, the 6dFGS optical spectrum only shows weak narrow H$\alpha$ and [O~III] and no broad lines, arguing against a classification as a Flat Spectrum Radio Quasar (FSRQ). The equivalent widths (EWs) for the H$\alpha$ and [O~III] lines are 1.6$\pm$0.4 \AA\ and 1.7$\pm$0.4 \AA, respectively. They are lower than 5\AA, which is the typical limit set to identify BL~Lac objects \citep{stickel91}. However, note that the combination of a low sensitivity and an intrinsically weak line-emitting AGN, diluted by a dominant galaxy continuum, such as FR~Is \citep{buttiglione09}, can equally lead to small EWs. 

Since the 6dFGS spectrum is not calibrated, we perform a qualitative photometric flux calibration, similar to the approach used by \citet{grandi16} to calibrate the 6dFGS spectrum of Tol~1326$-$379. Such method was already tested on a group of seven early-type emission-line galaxies in common between the 6dFGS and the SDSS surveys. To normalize the spectrum, we use the J-band image from the Two Micron All Sky Survey (2MASS, \citealt{skrutskie06}). After degrading the J-band image to match the seeing of the 6dFGS survey ($\sim$5.7$\arcsec$), we extract the flux from the image from an aperture of 6.7$\arcsec$, consistent with the diameter of the 6dFGS fiber where the spectrum comes from, and we measure J=12.12$\pm$0.2. To convert the flux into optical wavelenghts, we adopt a V$-$J = 2.43 typical of early-type galaxies \citep{mannucci01}. Assuming that the derived V-band magnitude is consistent with the spectrum continuum flux in the optical band of our object, we calculate from the EW the emission line fluxes, 1.3$\times$10$^{-14}$ and 8.2 $\times$10$^{-15}$ erg s$^{-1}$ cm$^{-2}$, corresponding to luminosities of 2.2$\times$10$^{40}$ and 1.4$\times$10$^{40}$ erg s$^{-1}$ for H$\alpha$ and [O~III], respectively. The tentative flux calibration of the spectrum yields to an uncertainty of a factor of 4 on the calculated fluxes and luminosities.

We can infer the BH mass, M$_{\rm BH}$, of the target from its K-band magnitude using the empirical relation found by \citet{marconi03} between the near-infrared bulge luminosity and the M$_{\rm BH}$. We use the K-band 2MASS total magnitude (9.55$\pm$0.03) to calculate its luminosity (4.35$\times$10$^{11}$ L$_{\odot}$). The BH mass estimated within a factor of $\sim$3 is 1.1$\times$10$^{9}$ M$_{\odot}$. 

Taking advantage of the [O~III] and M$_{BH}$ measurements, we can also infer other information on the AGN properties of IC~1531. Its bolometric luminosity L$_{\rm bol}$ corresponds to 4.8$\times$10$^{43}$ erg s$^{-1}$ by using the relation L$_{\rm bol}$=3500$\times$L$_{[\rm O~III]}$ \citep{heckman04} valid for low-luminosity radio AGN with similar luminosity of IC~1531. This value, along with the BH mass, leads to the Eddington luminosity ratio, L$_{\rm bol}$/L$_{\rm Edd}$, of 3.5$\times$10$^{-4}$ (within an error of a factor $\sim$5), indicating a low-luminosity accretion disc, typical of LEG \citep{best12}.

\section{Classification}\label{class}

In the 3LAC \citep{Acker} \ic~ is defined as blazar of uncertain type (BCUI) as the optical spectrum is considered not sensitive enough to discriminate between FSRQ and BL Lac. Our detailed study allows a more accurate classification.

The radio power, $P_{1,4\,GHz}=5.47\times 10^{23}\,W\,Hz^{-1}$, and the absolute magnitude in the R-band, $M_{R}=-23.37$, locate \ic~ in the region mainly occupied by FR I radio galaxies in the radio luminosity versus optical luminosity plane of \cite{Led96}.
Moreover, its NVSS radio map (Fig. \ref{NVSS}) shows a lobe-like south-eastern structure of $\sim$220~kpc, suggesting a misaligned jet. At a viewing angle of $\sim$5\dg, the corresponding deprojected  total linear size, assuming source symmetry, would exceed 5 Mpc.
Incidentally, we note that low power radio galaxies similar to \ic~ do not seem to reach Mpc extension (see \citealt{capetti17a}, \citealt{capetti17b}), that is preferentially observed in powerful radio sources (see 3CR and 2Jy sources, \citealt{is}, \citealt{singal}). 

In the color-color {\it WISE} diagram \citep{Wright}, its magnitudes ($w1-w2=0.031$ and $w2-w3=1.288$) locate \ic~  in the elliptical galaxy zone, indicating that the non-thermal emission of the source is not the dominant contribution at wavelengths 3.4--12~$\mu$m. The stellar population is also dominant in the optical spectrum (see Sec. \ref{opt}), although the [OIII] and H$\alpha$ lines are detected.
In principle, it is difficult to distinguish between a low-luminosity BL Lac object and a misdirected radio galaxy in the optical band when the AGN is overwhelmed by the galaxy emission.
We then took advantage of the diagnostic plane proposed by \citep{torresi18} to investigate this issue. They showed that the ratio between the [OIII] line and the 2--10~keV luminosity (R$_{\rm[OIII]}$) is a useful tool to separate misaligned and aligned jets.
As the optical emission line is isotropically emitted by circumnuclear gas, while the X-ray photons are emitted by the jet, R$_{\rm[OIII]}$ is expected to increase  with the jet inclination (i.e. when the Doppler boosting becomes less important and the X-ray luminosity less amplified). 
The average value measured for FR~Is\footnote{The sample is composed by 35 FR~Is belonging to the 3CR and 3CRR catalogs. The average value is estimated by using the Kaplan-Meyer (KM) estimator in ASURV including upper limits; \cite{lavalley92}.} is <R$_{\rm[OIII]}$>=-1.7$\pm$0.2 \citep{torresi18}, while BL Lacs have generally lower values <R$_{\rm[OIII]}$>=-3.3$\pm$0.2 in agreement with the Doppler boosting of the X-ray radiation. The estimated ratio for IC~1531 is R$_{\rm[OIII]}$=-1.4$\pm$0.9, similarly to FR~Is, suggesting that our source shares the same optical-X-ray properties of FR I radio galaxies.

\begin{figure}
	\includegraphics[width=\columnwidth]{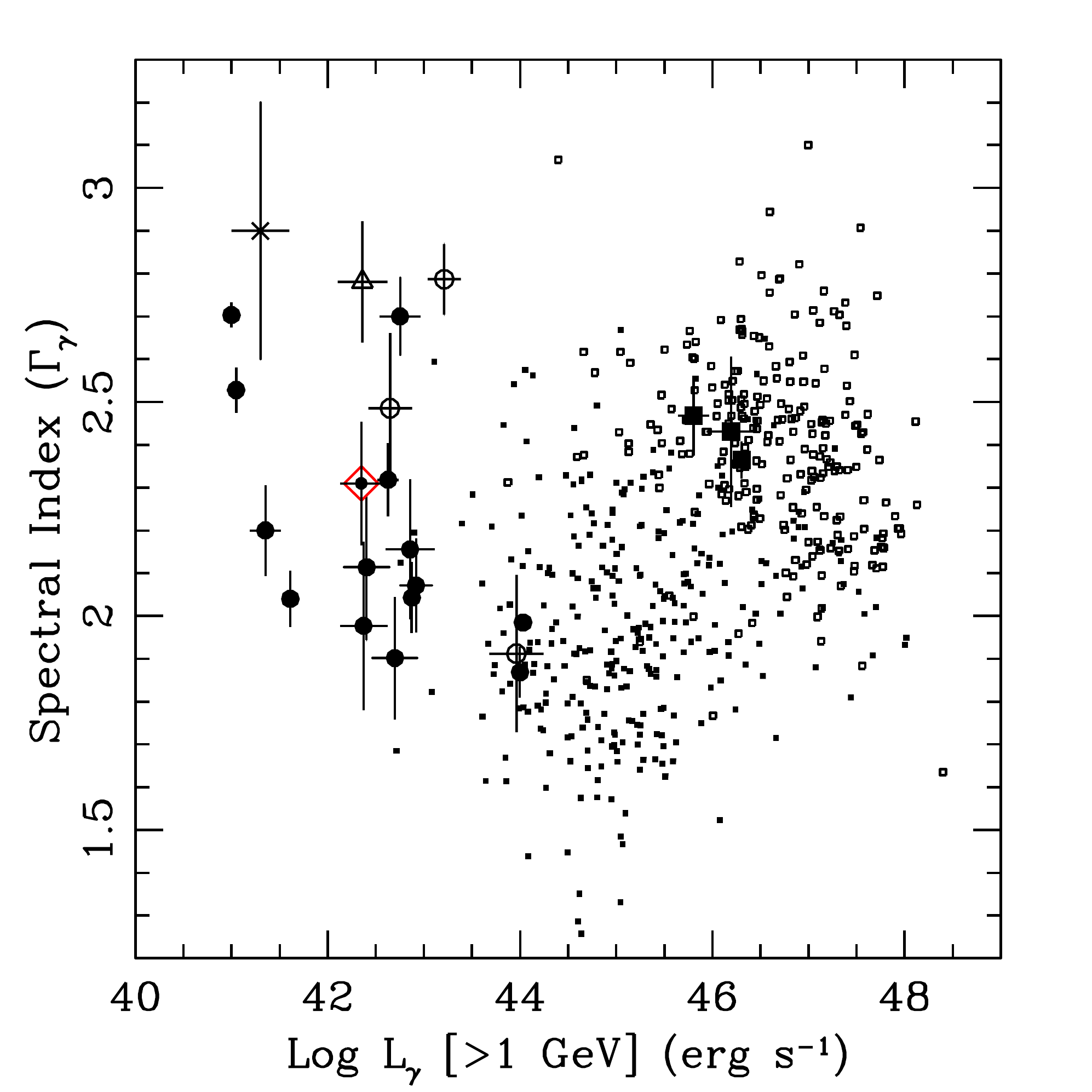}
    \caption{Gamma-ray spectral slope, $\Gamma_\gamma$, versus 1-100 GeV luminosity. IC~1531 (red diamond) is located in the 3LAC MAGN (FRIs, FRIIs; filled and empty circles respectively, and SSRQs; big filled squares) region and outside the Blazar (BL LAcs, FSRQs; small filled and empty squares, respectively) zone. The FR0 radio galaxy TOL~1326-379 \citep[empty triangle, ][]{grandi16} and the Compact Symmetric Object PKS1718-649 \citep[black cross, ][]{Mig16}  are also shown.}
    \label{gamma}
\end{figure}

Finally we note that the source is a low luminosity  \g-ray emitter ($Log\,L_{\gamma\,>1\,GeV\, [erg\,s^{-1}]}=42.3\pm 0.2$) with a steep spectrum ($\Gamma_{\gamma}=$2.3$\pm$0.1). In the  diagnostic  $\Gamma_{\gamma}-L_{\gamma\,[>1\,GeV]}$ plane \citep{MAGN}, the different class of AGN lie in different parts of the diagram. FSRQs and \textit{Steep Spectrum Radio Quasars (SSRQs)} occupy the right side of the plane at higher gamma luminosity and steep spectra, while the BL Lacs are located in the bottom right side of the diagram, having lower luminosity and flatter spectra. The majority of the radio galaxies are arranged at low luminosities outside the blazar strip\footnote{ We note that a number of MAGNs are found in the region occupied by blazars (see Figure \ref{gamma}). Indeed, these targets display a blazar-like, flaring behavior \citep[see, for example PKS 0521-36,][]{Dam15} and it is plausible that their small-scale jets are more closely aligned with the observer's line of sight. 
Similarly, there is not a clear boundary between BL Lacs and FSRQs. One can find transitional sources, which have intermediate properties between high- and low-power radio sources \citep[][]{Sam07}. In addition, it should be noted that a FSRQ caught at the time of a flare can be misclassified as a BL Lac \citep{Ghi11}.}. \ic~falls exactly in the zone populated by radio galaxies (Figure \ref{gamma}).

On the basis of the multi-band properties, we conclude that \ic~ is  a  FR I radio galaxy showing a moderate Doppler boosted flux amplification.

\section{Nuclear SED}\label{sed}
\begin{figure}
	\includegraphics[width=\columnwidth]{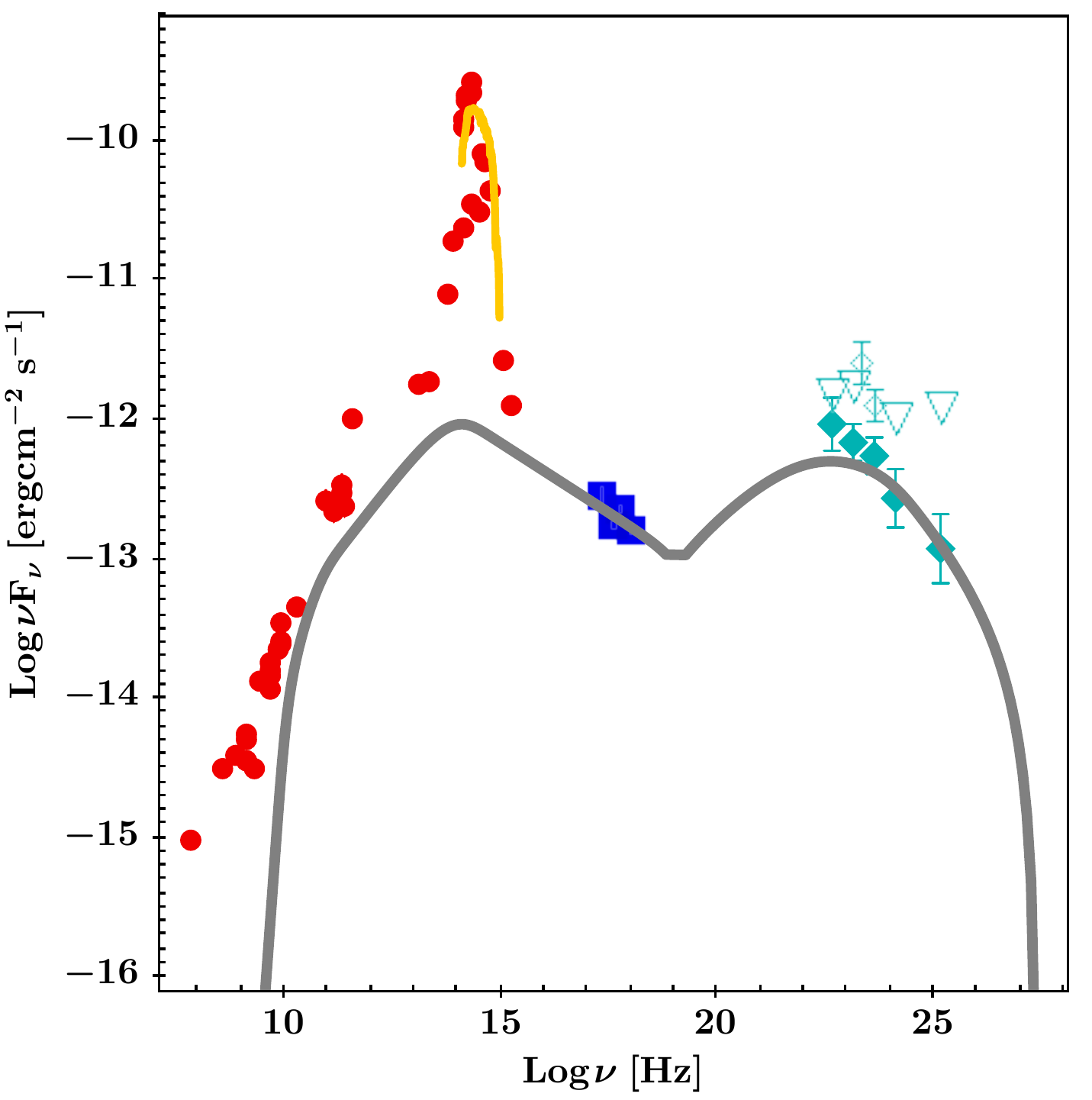}
    \caption{The observed nuclear SED of IC~1531 is described by a leptonic, one-zone synchrotron-SSC model, assuming an observer's viewing angle $\theta=$15\dg (gray thick line). The model parameters are reported in Table \ref{tab_model}. The filled red points are the data collected from the literature and on-line archives (see Table \ref{sed}), the filled diamonds are the \g-ray fluxes from the 3FGL, the empty diamonds are from the 2FGL and the empty triangles are the upper limits. The filled squares are the \Cha~fluxes of the core. The orange line is the galaxy template \citep{mannucci01}.}
    \label{SED}
\end{figure}

We assembled the broadband SED of the core of \ic\ by combining the analyzed data with those available in literature and in on-line databases (see Table \ref{sed}).
Although the data are not simultaneous, we note that the multi-epoch observations in the radio and X-ray bands do not indicate significant variability. Similarly, the 3FGL reports a variability index of 42.5, which indicates a low probability for the source to be variable in the \g-ray band.
The observed nuclear SED of \ic~is shown in Figure \ref{SED}. The emission of the host galaxy dominates the SED in the optical-IR band. 
There is no evidence of a contribution of the disc continuum emission, confirming that the AGN is powered by a low-luminosity radiatively inefficient accretion disc (see also Sec. \ref{opt}). The upper limit to the disc's UV luminosity derived from the observations ($L_{UV}\lesssim 3.0 \times 10^{42}$ \ergs, see Table \ref{sed}) is significantly  smaller than the AGN bolometric luminosity derived in Section 5.

The compact jet produces radio to \g-ray non-thermal emission.
We modeled the jet's SED in the framework of leptonic, one-zone models.
The emission is produced by a spherical blob of radius $R_{blob}$, moving with a bulk Lorentz factor $\Gamma_{\rm bulk}$. 
We assumed a conical jet and set $R_{blob}=\psi z_{dist}$, with $z_{dist}$ and $\psi$ being the distance from the base of the jet and the jet's semi-opening angle, respectively, and we fixed $\psi=$0.1 rad \citep[see e.g.][for details]{Mig14}. 
Interacting with the jet's magnetic field ($B$), the electrons in the blob radiate synchrotron photons, which are up-scattered to higher energies \citep[synchrotron self-Compton, SSC][]{Mar92}. 
Given the limits on the disc luminosity and emission lines, Comptonization of external photons fields \citep[including disc's reprocessed emission in the broad line regions and torus, ][]{Sik94} is not a significant contribution and was not considered in the modeling.\\
\indent
Constraints to some of the model parameters could be derived from the observations \citep[see e.g.,][]{Tavecchio98,Ghisellini13}. 
The features of the observed SED can be used to define the shape of the energy distribution of the radiating electrons $N(\gamma)$.
The synchrotron spectrum peaks at approximately $\nu_{syn}\sim$10$^{13-15}$ Hz and extends to the X-ray band, as the steep X-ray spectral index ($\alpha_X=$1.2\er 0.1) indicates. In the high-energy band, the SSC peak, $\nu_{SSC}$, can be located at $\approx$10$^{22-24}$ Hz. The observed SED suggests that the fluxes of the two peaks are similar, F$_{\rm syn/SSC,peak}\lesssim$ 10$^{-12}$ \ergcm. 
We set the synchrotron spectral index at energies beyond the peak equal to $\alpha_{syn,2}=\alpha_X$, a value that is also consistent with the spectral index measured by \fermi-LAT, and we assumed a slope $\alpha_{syn,1}\sim$0.6 below the peak. 
Based on these features, we assumed a broken power-law shape for $N(\gamma)$. The energy break $\gamma_b=(3\nu_{SSC}/4\nu_{syn})^{1/2}$ falls in the range $\approx 10^{3-4}$ and the spectral indexes below and above $\gamma_b$ are $p_1\equiv 2\alpha_{syn,1}+1=2.2$ and $p_2\equiv 2\alpha_{syn,2}+1=3.4$, respectively.\\
\indent
The radio observations provided us with constraints on the inclination of the jet with respect to the observer's line of sight, $\theta$. 
A maximum $\theta$ of $\sim$30\dg~was inferred by using the relation between the core radio power at 5 GHz and the total radio power at 408 MHz in \citet{Giovannini01}.
Based on the jet-counterjet radio flux ratio and keeping into account the uncertainties on the measurements (see Appendix B for the details), we derived a range of $\theta$ between $\sim$10\dg~and $\sim$20\dg. 
We assumed $\theta=$15\dg~as fiducial value and investigated the implications of smaller/larger values for the jet structure and energetics.
The modeled non-thermal beamed SED of the jet for $\theta=$15\dg~is shown in Figure \ref{SED}. The values of the model parameters are reported in Table \ref{tab_model} together with the jet's Poynting flux ($L_B$), the kinetic power ($P_{jet}=L_e+L_p$, with $L_e$ and $L_p$ powers in radiating electrons and cold protons, respectively) and the radiating power (L$_{rad}$). We note that the inferred jet powers are comparable with the bolometric AGN luminosity derived from [O~III] luminosity, confirming that the SED of this object is jet-dominated.
Given the small dimensions and the value of the $B$ field of the emitting region, the modeled synchrotron spectrum is self-absorbed at $\lesssim$10$^{11}$ Hz and does not account for the radio emission at $<$GHz frequencies, which is assumed to come from larger regions of the jet. Similarly, at frequencies below the synchrotron peak the modeled jet emission remains below the data points, as the $WISE$ data indicate that the host galaxy is still dominant at $10^{13}$ Hz (see Sec. \ref{class}).

\begin{table*}
\begin{center}
\begin{tabular}{ccccccccccc}
\hline
\hline 
$\theta$ & $\Gamma_{\rm bulk}$ & $\gamma_{min}$ & $\gamma_{max}$ & $\gamma_{b}$ & $U^{'}_{e}/U^{'}_{B}$ & $L_{B}$ & $L_{e}$ & $L_{p}$ & $P_{jet}$ & $L_{rad}$\\
(1)   & (2)   &(3)   &(4)   &(5) &(6)  &(7)  &(8)  &(9)  &(10)  &(11)\\
\hline
\multicolumn{11}{c}{\bf radio galaxy}\\
12  &4.5  &180 &$5\times 10^{6}$  &$9\times 10^{3}$  &$8.7\times 10^{-3}$/$2.5\times 10^{-5}$  &41.23  &43.46  &43.90 &44.04 &41.94\\
15  &4     &180 &$5\times 10^{6}$ &$9\times 10^{3}$  &$7.0\times 10^{-3}$/$1.4\times 10^{-4}$   &41.72  &43.11  &43.56 &43.69 &42.15\\
20 &2   &180 &$5\times 10^{6}$  &$7.5\times 10^{3}$  &$9.5\times 10^{-3}$/$2.6\times 10^{-4}$  &41.49  &42.75  &43.18 &43.33 &42.18\\
\multicolumn{11}{c}{\bf blazar}\\
5 &10 &180 &$5\times 10^{6}$ &$8.0\times 10^{3}$ &$5.4\times 10^{-3}$/$2.6\times 10^{-6}$  &40.78  &43.80 &44.25 &44.38 &41.02\\
\hline
\tabularnewline
\end{tabular}
\end{center}
\caption{Model parameters for the jet nuclear SED of \ic. Columns: 1- jet viewing angle in degrees; 2- bulk motion of the emitting region; 3,4,5- minimum, maximum and break random Lorentz factors of the radiating electrons (a broken power-law shape of the energy distribution is assumed); 6- energy densities of the electrons/magnetic field in the blob comoving frame in units of ergs cm$^{-3}$; 7- jet Poynting flux; 8- power in bulk motion of the radiating electrons; 9- power in bulk motion of  cold protons (one cold proton per radiating electron is assumed); 10- jet kinetic power ($L_e+L_p$); 11- jet radiative power. Quantities in 7-11 are in logarithmic scale (in units of erg s$^{-1}$).}
\label{tab_model}
\end{table*}

\section{Discussion}
The multi-wavelength study of \ic~presented in the previous sections favors a classification of the source as a radio galaxy. The radio and X-ray brightness of the large-scale radio structure is decreasing moving out of the core, and the X-ray spectrum of the kpc jet is compatible with being synchrotron emission, as typical for low-power, FR I radio galaxies \citep[][for a review]{Jet}. 

\ic~shares the  properties of the LAT detected FR Is \citep[see][and references therein]{torresi18}. These  sources are continuously detected by LAT and show low-level or no variability in \g-rays \citep{fermi,Sah18}. The notable exception is NGC 1275 which couples a long-term continuous brightening with short-term (down to few days) flaring activity \citep{Tan18}. However, this radio source is peculiar in that its radio activity has recently re-started and shows strong evolution of its small-scale radio structure \citep{Nag10}. Differently, FR IIs are detected only when they undergo rapid and intense flares, which appear to be accompanied by the ejection of a new component \citep{Gra12,Cas15,tanaka}.
As for IC 1531, the nuclear X-ray emission of LAT FRIs is non-thermal. 
The X-ray emission of the broad line radio galaxies (BLRGs) observed by \fermi~is instead dominated by accretion-related processes \citep{GP07,Kat11}. Indeed, FRIIs and some BLRGs, such as 3C 120, have bright disc emission and, likewise FSRQs, there could be a contribution from external Compton on the photons from the disc (or reprocessed in the broad line regions and torus).
 
In \ic, the radio data suggest a $\theta$ ranging from $\sim$10\dg~to $\sim$20\dg. Within these limits, we were able to explain the nuclear SED  with a synchrotron and SSC model by assuming a moderate/low bulk motion $\Gamma_{\rm bulk}\sim$2--4.5. Such values are in agreement with those inferred by modeling the SED of other MAGNs detected by \fermi, while BL Lac sources have typically $\Gamma_{\rm bulk}\sim$10--20.

In the framework of the unified model of jetted AGNs \citep{Urry-Padovani,Pad17}, radio galaxies are the misaligned counterparts of blazars and the jets of the two classes should have the same structure. The hypothesis to reconcile the different $\Gamma_{\rm bulk}$  is that, at increasing viewing angles, the de-boosting of the emission from the blazar region makes visible the radiative components of mildly relativistic regions \citep{Chiaberge}. Indeed, the Doppler factors of these slower regions, $\delta\sim$2--7, are smaller than in BL Lac objects ($\delta\gtrsim$10), limiting the factor of amplification of the intrinsic luminosities, $L'$, in the observer reference frame ($L=\delta^4 L'$). Hence the observed \g-ray fluxes of MAGNs are relatively faint and require sensitive observations to be detected.

The sensitivity reached in ten years of LAT sky survey opens at the possibility to realize a tomography of the regions of the jet where the most energetic emission is produced.
In the case of \ic, for simplicity we considered a dynamical structure that is orthogonal to the jet main axis (i.e. the emitting region is decelerating while moving away from the core). However, observations indicate that a stratified geometry, with a fast spine surrounded by a slow layer in its simplest configuration, is equally possible. Interestingly, in a handful of sources, imaging in the radio to sub-mm band with very-long-baseline interferometry (VLBI) have probed AGN jets down to hundreds of gravitational radii from the central BH and revealed a limb-brightened transversal structure \citep[see Mrk~501, M~87, Cygnus~A and 3C~84;][]{Gir04,Had16,Boc16,Nag14,Gio18}. These results suggest that mildly relativistic, emitting regions may be present already at sub-pc scales.  Given the limited sample, it is not yet clear whether all jets, independently from their power, develop the same structure at small scales and how such a structure relates to the morphological differences which are seen at kiloparsec scales \citep[FR I and FR II types, see for example the case of PKS~1127$-$145;][]{Sie07}.
Indeed, because of the long timescales of the radio duty cycles in AGNs \citep[$\gtrsim$10$^7$ yrs][and references therein]{Sha08,Bri17}, it is not easy to determine the factor(s) shaping the morphology of the giant radio galaxies. The role of the environment where the radio source forms, as well as the connection with the different accretion modes of the AGN are currently under investigation \citep[see e.g.][]{Led96,Bes09,MB17,Cro18}. 
Because of BH mass scaling relations, precious insights on the nature of the accretion and ejection phenomena come from X-ray binaries hosting stellar mass black holes, whose radio outbursts are typically observable on human timescales \citep[see ][ for a review]{FG14}.

The acceleration mechanisms and radiative processes active in the jet's layer could be different from those in the highly relativistic spine. For example, continuous acceleration of ultra-relativistic electrons in turbulent boundary layers has been proposed to explain the X-ray synchrotron emission of FR I jets at kiloparsec scales \citep[][and references therein]{SO02}, whereas strong shocks and magnetic reconnection are candidate processes in the spine \citep[see e.g.][]{Gia13,Sir15,Nal16}. 
In \ic, a SSC process can account for the observed emission and returns rather standard jet parameters and powers. However, in other MAGNs \citep[e.g. 3C~84, NGC~6251][]{Tav14,Giulia}, a SSC model implies extreme jet parameters (in terms of e.g. pressure, magnetization..), which would challenge the jet stability, or very large powers. On the other hand, the synchrotron emission produced by the spine(/layer) represents a natural source of seed photons for the electrons in the layer(/spine). 
Models of jets with multiple, radiatively interacting regions have been proved effective in reproducing the GeV and even TeV emission of some of the MAGN \citep{GK03,spine-layer}.

To conclude, although disfavored by the observations, we briefly discuss the case of an aligned jet (say $\theta=$5\dg). Remarkably, this would make \ic~the BL Lac in the 3FGL with the lowest known redshift. Its radio luminosity ($L_r\sim$10$^{39.9}$ erg s$^{-1}$) places \ic~in the lower radio-luminosity boundary of the class.
This is a relatively unexplored regime because the identification of intrinsically faint BL Lac objects can be challenging \citep[see][for a discussion]{CR15}, and yet extremely important in order to build their luminosity function (LF). By using their developed selection method based on mid-infrared emission, \citet{CR15} argued in favor of a break  in the local BL Lac LF at $L_r\sim$10$^{40.6}$ erg s$^{-1}$, which could be possibly related to the power necessary to launch a jet. In the BL Lac scenario, \ic~would fall below this break. Detections of intrinsically faint \g-ray BL Lacs at low redshift are key to define the local \g-ray LF of this class and investigate a possible cosmological evolution, allowing also to estimate their contribution to the isotropic \g-ray background \citep{Aje14}.

\section{Summary \& Conclusions}
In this paper we performed a multi-wavelength study of the radio source \ic, associated with the \g-ray source 3FGL~J0009$-$3206. 
Its large-scale radio and X-ray morphologies and different diagnostic methods in the optical to infrared bands suggest that the source is likely a FR I radio galaxy whose jet is seen at moderate inclination angles ($\theta\sim 15$\dg). As such, we propose \ic~ as a member of the MAGNs detected by LAT. 
The \g-ray flux can be explained in terms of SSC emission from a region with $\Gamma_{\rm bulk}\sim$4. In the framework of the unified scheme of AGNs, this supports the presence of multiple emitting regions with different velocities, such could be a spine-layer structure. Intriguingly, the processes which accelerate the particles radiating at high energies could be different with respect to those in the highly relativistic regions of blazars, possibly explaining some of the  observational differences between blazars and MAGNs (such as e.g. the lack of significant variability and flares in X-to-\g-rays). High-energy and multi-wavelength studies of a larger number of sources at the boundaries between radio galaxies and blazars have the potential to provide us with a tomography of the jet regions where the most energetic emission is produced. The enhanced sensitivity reached by \fermi~and the advent of large-field, sky-surveying observatories such as CTA, eROSITA, LSST and SKA will be key to make progress in this research.

\section*{Acknowledgements}
The research leading to these results has received funding from the European Union\'s Horizon 2020 Programme under AHEAD project (grant agreement n. 654215). Partial support to this work was provided by the NASA grant AR4-15009X. G.M. acknowledges the financial support from the UnivEarthS Labex program of Sorbonne Paris Cit\'e (ANR10LABX0023 and ANR11IDEX000502).
We thank the anonymous referee for comments and suggestions which helped us to improve the manuscript.
T.B. acknowledges financial contribution from the agreement ASI-INAF n.2017-14-H.0.
MAPT acknowledges support from the Spanish MINECO through grants AYA2012-38491-C02-02 and AYA2015-63939- C2-1-P.  A.S. was supported by NASA contract NAS8-03060 (Chandra X-ray Center).
E.T. acknowledges financial support from ASI-INAF grant 2015- 023-R.O.
We acknowledge financial support from PRIN INAF 2016.
This work made use of data supplied by the UK Swift Science Data Centre at the
University of Leicester, and archival data from the VLA. The National Radio Astronomy Observatory is a facility of the National Science Foundation operated under cooperative agreement by Associated Universities, Inc.




\bibliographystyle{mnras}

\bibliography{bibliografia}



\appendix

\section{Some extra material}
In Table \ref{sed}, we report the informations relative to the flux measurements in the radio to \g-ray band used to compile the SED of \ic. 

\begin{table*}
\begin{center} 
\footnotesize
\begin{tabular}{cccc}
\hline\hline
\tabularnewline
Frequency&Flux&Angular&Reference\\    
  (Hz)  &(\lumunits)& resolution   \\ 
\tabularnewline
\hline
\tabularnewline
$7.4\times 10^{7}$	&$(9.3\pm 1.1)\times 10^{-16}$	&80"&\textit{VLSS}\tabularnewline
$4.08\times 10^{8}$	&$(3.1\pm 0.2)\times 10^{-15}$	&~3'&\citet{Large}\tabularnewline
$8.43\times 10^{8}$	&$(3.8\pm 0.1)\times 10^{-15}$	&-&\citet{SUMSS}\tabularnewline
$1.4\times 10^{9}$	&$(5.4\pm 0.2)\times 10^{-15}$	&45"&\textit{NVSS}\tabularnewline
$1.47\times 10^{9}$	&$ 3.4\times 10^{-15}$	&1"&\citet{vanGorkom}	\tabularnewline
$2.3\times 10^{9}$	&$3.0\times 10^{-15}$	&90mas&\citet{Slee}	\tabularnewline
$2.7\times 10^{9}$	&$(13.0\pm 0.5)\times 10^{-15}$	&8'&\citet{Shimmins}\tabularnewline
$4.8\times 10^{9}$	&$1.4\times 10^{-14}$	&5"&'' 	\tabularnewline
$4.9\times 10^{9}$	&$1.1\times 10^{-14}$	&VLA&'' \tabularnewline
$5\times 10^{9}$	&$(15.7\pm 0.8)\times 10^{-15}$		&10.8"x10.8"&\citet{Murphy}\tabularnewline
$5\times 10^{9}$	&$ (17.5\pm 1)\times 10^{-15}$	&8'&''\tabularnewline
$8\times 10^{9}$	&$(2.2\pm 0.1)\times 10^{-14}$		&10.8"x10.8"&''\tabularnewline
$8.4\times 10^{9}$	&$2.4\times 10^{-14}$	&27mas&'' 	\tabularnewline
$2\times 10^{10}$	&$(4.4\pm 0.2)\times 10^{-14}$	&10.8"x10.8"&''\tabularnewline
$1.499\times 10^{11}$	&$(2.2\pm 0.4)\times 10^{-13}$	&27.5"&\citet{Knapp}\tabularnewline
$2.306\times 10^{11}$	&$(2.9\pm 0.6)\times 10^{-13}$	&19.5"&''\tabularnewline
$2.725\times 10^{11}$	&$(2.4\pm 0.4)\times 10^{-13}$	&18.5"&''\tabularnewline
$2.998\times 10^{12}$	&$(1.4\pm 0.5)\times 10^{-11}$	&3.5'&\citet{Knappaltro}\tabularnewline
$3.747\times 10^{11}$	&$(9.9\pm 1.6)\times 10^{-13}$	&17"&''\tabularnewline
\tabularnewline

$1\times 10^{11}$	&$(2.6\pm 0.5)\times 10^{-13}$&&\textit{PR2-2015 (Planck Legacy Archive)}\tabularnewline
$2.17\times 10^{11}$	&$(3.3\pm 0.6)\times 10^{-13}$&&''\tabularnewline
\tabularnewline

$1.363\times 10^{13}$	&$(17.9\pm 0.2)\times 10^{-13}$&12.0''&\textit{WISE}\tabularnewline
$2.498\times 10^{13}$	&$(18.2\pm 0.5)\times 10^{-13}$&6.5''&''\tabularnewline
$6.517\times 10^{13}$	&$(7.8\pm 0.2)\times 10^{-12}$&6.4''&''\tabularnewline
$8.817\times 10^{13}$	&$(18.6\pm 0.4)\times 10^{-12}$&6.1''&''\tabularnewline
\tabularnewline

$1.388\times 10^{14}$	&$(12.2\pm 0.4)\times 10^{-11}$&2''&\textit{2MASS}\tabularnewline
$1.388\times 10^{14}$	&$(14.1\pm 0.4)\times 10^{-11}$&2''&''\tabularnewline
$1.806\times 10^{14}$	&$(18.8\pm 0.5)\times 10^{-11}$&2''&''\tabularnewline
$1.806\times 10^{14}$	&$(20.9\pm 0.6)\times 10^{-11}$&2''&''\tabularnewline
$2.418\times 10^{14}$	&$(22.3\pm 0.4)\times 10^{-11}$&2''&''\tabularnewline
$2.418\times 10^{14}$	&$(26.0\pm 0.5)\times 10^{-11}$&2''&''\tabularnewline
\tabularnewline

$3.893\times 10^{14}$	&$(8.0\pm 0.3)\times 10^{-11}$&-&\textit{AAVSO}\tabularnewline
$4.797\times 10^{14}$	&$(6.9\pm 0.2)\times 10^{-11}$&-&''\tabularnewline
$6.246\times 10^{14}$	&$(4.31\pm 0.05)\times 10^{-11}$&-&''\tabularnewline
\tabularnewline

$1.292\times 10^{15}$	&$(2.6\pm 0.2)\times 10^{-12}$	&6''&\textit{GALEX}\tabularnewline
$1.947\times 10^{15}$	&$(1.2\pm 0.2)\times 10^{-12}$	&4''&''\tabularnewline
\tabularnewline

$4.84\times 10^{22}$	&$<1.6\times 10^{-12}$	&&Fermi2FglLC	\tabularnewline
$4.84\times 10^{22}$	&$(9.0\pm 4.9)\times 10^{-13}$	&&Fermi3FGL\tabularnewline
$1.452\times 10^{23}$	&$<1.9\times 10^{-12}$	&&Fermi2FglLC	\tabularnewline
$1.452\times 10^{23}$	&$(6.6\pm 2.4)\times 10^{-13}$	&&Fermi3FGL\tabularnewline
$2.42\times 10^{23}$	&$(2.5\pm 1.1)\times 10^{-12}$	&&Fermi2FglLC\tabularnewline
$4.84\times 10^{23}$	&$(1.3\pm 0.4)\times 10^{-12}$	&&Fermi2FglLC\tabularnewline
$4.84\times 10^{23}$	&$(5.3\pm 1.9)\times 10^{-13}$	&&Fermi3FGL\tabularnewline
$1.452\times 10^{24}$	&$<1.1\times 10^{-12}$	&&Fermi2FglLC	\tabularnewline
$1.452\times 10^{24}$	&$(2.7\pm 1.7)\times 10^{-13}$	&&Fermi3FGL\tabularnewline
$1.452\times 10^{25}$	&$<1.3\times 10^{-12}$	&&Fermi2FglLC	\tabularnewline
$1.452\times 10^{25}$	&$(1.2\pm 0.9)\times 10^{-13}$	&&Fermi3FGL\tabularnewline
\tabularnewline
\hline
\end{tabular}
\end{center}
\caption{ Multi-wavelength data of \ic~taken from the literature and on-line databases. Columns: 1- observed frequency; 2- flux and error (when available); 3- angular resolution of the instrument; 4- references and mission database.
}
\label{sed} 
\end{table*}

\section{Constraints}
In this section, we detail the procedure that we followed to derive constraints on the jet's inclination angle and speed.\\
The flux density ratio of the approaching to the receding component, $S_a$ and $S_r$, can be written as follows \citep{MR99}:

$$ R \equiv {S_a \over S_r} = \biggl({1 + \beta \cos~ \theta
\over 1 - \beta \cos~ \theta} \biggr)^{k-\alpha} $$

where ${S_a \over S_o} = \delta_a^{k-\alpha}$, ${S_r \over S_o} = \delta_r^{k-\alpha} $,
are the ratios of observed to emitted flux densities of the approaching and receding components, $\beta = v/c$ is the intrinsic jet speed, in units of $c$, 
 $\theta$ is the angle of the jet to the line of sight, 
$\alpha$ is the spectral index of the emission
($S_\nu \propto \nu^\alpha)$, and $k$ is a parameter that
accounts for the geometry of the ejecta, with k = 2 for a continuous jet
and k = 3 for discrete condensations.

The (true) jet speed $\beta$ and the bulk Lorentz factor, $\Gamma_{\rm bulk}$, are related as follows: $\Gamma_{\rm bulk} = (1 - \beta^2)^{-1/2}$, so that  $\beta = (1 - \Gamma^{-2})^{1/2}$.
Although we lack a precise knowledge of the values of $\beta$ and/or $\theta$, we know that for the angle $\theta_m$ that maximizes the value of $\beta$ the following equality applies: 
$\cos \theta_m = \beta$.
Hence, 

$\cos~\theta_m = \beta = (1 - \Gamma_{\rm bulk}^{-2})^{1/2}$,

and we get

$$
{1 + \beta \cos~ \theta_m \over 1 - \beta \cos~ \theta_m} = 
{1 + \beta^2 \over 1 - \beta^2} = 
{1 + (1 - \Gamma_{\rm bulk}^{-2}) \over \Gamma_{\rm bulk}^{-2}} = 
2\, \Gamma_{\rm bulk}^2 - 1
$$.

Substituting in the Equation for $R$ and rearranging, we get the angle $\theta_m$ that maximizes $\beta$ (and therefore $\Gamma_{\rm bulk}$), as a function of the flux density ratio, $R$:

$$
\theta_m = \arccos \biggl( {R^m - 1 \over R^m + 1}\biggr)^{1/2}
$$,

where $m = 1/(k - \alpha)$. We assume $k = 2,3$, no significant variability between the L-band and X-band observations, and flux uncertainties within 10\%. For the nominal flux values, we obtain $\theta_m=$14.8\dg, $\beta=$0.967 and $\Gamma_{\rm bulk} =$3.93, with a range of values for the angle $\theta_m$ spanning between 10\dg~and 20\dg.  


\bsp	
\label{lastpage}
\end{document}